\documentclass[aps,prl,reprint,superscriptaddress]{revtex4-1}
\usepackage{amsmath,amssymb,graphicx,epspdfconversion,hyperref,cleveref}

\begin{document}

\title{Entanglement Devised Barren Plateau Mitigation}%

\author{Taylor L. Patti}
\email{taylorpatti@g.harvard.edu}
\affiliation{Department of Physics, Harvard University, Cambridge, Massachusetts 02138, USA}
\author{Khadijeh Najafi}
\affiliation{Department of Physics, Harvard University, Cambridge, Massachusetts 02138, USA}
\affiliation{IBM Quantum, IBM T.J. Watson Research Center, Yorktown Heights, NY 10598 USA}
\author{Xun Gao}
\affiliation{Department of Physics, Harvard University, Cambridge, Massachusetts 02138, USA}
\author{Susanne F. Yelin}
\affiliation{Department of Physics, Harvard University, Cambridge, Massachusetts 02138, USA}
\date{December 22nd, 2020}%

\begin{abstract}
Hybrid quantum-classical variational algorithms are one of the most propitious implementations of quantum computing on near-term devices, offering classical machine learning support to quantum scale solution spaces. However, numerous studies have demonstrated that the rate at which this space grows in qubit number could preclude learning in deep quantum circuits, a phenomenon known as barren plateaus. In this work, we implicate random entanglement as the source of barren plateaus and characterize them in terms of many-body entanglement dynamics, detailing their formation as a function of system size, circuit depth, and circuit connectivity. Using this comprehension of entanglement, we propose and demonstrate a number of barren plateau ameliorating techniques, including: initial partitioning of cost function and non-cost function registers, meta-learning of low-entanglement circuit initializations, selective inter-register interaction, entanglement regularization, the addition of Langevin noise, and rotation into preferred cost function eigenbases. We find that entanglement limiting, both automatic and engineered, is a hallmark of high-accuracy training, and emphasize that as learning is an iterative organization process while barren plateaus are a consequence of randomization, they are not necessarily unavoidable or inescapable. Our work forms both a theoretical characterization and a practical toolbox; first defining barren plateaus in terms of random entanglement and then employing this expertise to strategically combat them.
\end{abstract}

\maketitle

\section{Introduction}
The rapid development of noisy quantum devices \cite{Preskill2018quantumcomputingin} has lead to great interest in hybrid quantum-classical variational algorithms, through which classical machine learning techniques are employed to prepare, sample, and optimize states on noisy quantum hardware \cite{McClean_2016, Farhi2014, Peruzzo2014, Yung2014, Wecker2015}. Not only do these algorithms show potential for a variety of near-term applications \cite{Kandala2017}, they are inherently robust against certain coherent errors and are free to minimize decoherence effects through the exploration of unconventional gate sequences. Of particular interest are quantum neural networks (QNNs) \cite{Farhi2018}, in which quantum input states are transformed into output states by a parametrized quantum circuit (PQC). The output states then undergo a series of measurements, collectively referred to as a cost function, and the measurement results are used to optimize the circuit.

Although QNNs offer a straight-forward approach, their implementation can be quite challenging. Among the greatest of these difficulties are barren plateaus \cite{McClean2018}: regions of the cost function's parameter space where it is rather constant, varying too little for successful gradient-based optimization. While in shallow circuits these barren landscapes are cost function-dependent \cite{Cerezo2020, Uvarov2020}, the effect is cost function-independent for circuits that are sufficiently deep. Moreover, even gradient-free algorithms can be impacted \cite{Arrasmith2020, Anand2020}. While certain restricted subsets of PQCs are somewhat resilient to barren plateaus \cite{Sharma2020, Wiersema2020}, the most general implementation, known as the ``hardware efficient ansatz", becomes exponentially barren with increasing qubit number. Numerous techniques have been suggested for the amelioration of barren plateaus, including layerwise and symmetry based training \cite{Skolik2020, Fontana2020}, correlated and identity-esq circuit initialization \cite{Volkoff2020, Grant2019}, and quantum convolutional neural network protocols \cite{Pesah2020}, but they have yet to form a complete toolbox that is suitable for large-scale, general purpose QNNs.

Likewise, our understanding of barren plateaus is extensive yet far from complete. For some years, it has been understood that barren plateaus are a consequence of concentration of measure \cite{Bremner2009, Gross2009}, stemming from the effects of randomness on the exponential dimension of quantum state space.  More recently, the relationship between entanglement and barrenness has been explored by quantum scrambling studies \cite{Holmes2020} and in terms of visible and hidden units \cite{Marrero2020}. However, we still lack comprehensive understanding of how entanglement induces barren plateaus with respect to cost function register size, qubit connectivity, and circuit depth. As a result, barren plateau mitigation strategies that rely on these insights have yet to be developed.

In this work, we give a detailed account of how random entanglement leads to barren plateau formation. While barren plateaus can be both noise-independent and noise-induced \cite{Wang2020}, we here consider only the former variety. In particular, we derive the relationship between cost function barrenness and qubit entanglement, including the rate at which barrenness scales with circuit depth. As our findings quantify barrenness via the entanglement of specific qubit subsets, we develop partitioning methods that initially or continuously restrict such entanglement. This generates non-barren cost function landscapes and thus improves circuit learning. We find that initially partitioned circuits not only learn faster, but often produce less entangled solutions. As entangled states are more sensitive to decoherence, this factorizability can decrease the number of measurements required to accurately estimate the cost function, potentially reducing the problematic number of expectation values required for each circuit iteration \cite{Verteletskyi2020, Gokhale2019, Izmaylov2019}.

In order to verify and exploit these findings, we design a classical meta-learning protocol that avoids barren plateaus while generating an arbitrary circuit with rich entanglement structure. In contrast to other QNN meta-learning proposals \cite{Verdon2019}, which address specific problem classes, ours is suitable for general PQCs. Moreover, as our meta-learning technique does not pre-train circuit output, it is itself immune to barren plateaus. Furthermore, we model a real-time regularization process that penalizes forms of entanglement that are potentially problematic and show that this method ameliorates barren landscapes, decreasing both training time and error. We also make the novel identification of barren plateaus as a form of Langevin noise in the circuit parameter space and demonstrate the effectiveness of injecting additional Langevin noise into the training process, a technique that has been used to combat overfitting in deep classical neural networks \cite{Welling2011}. Finally, we draw a parallel between entanglement dynamics and the improved performance of QNNs in certain measurement bases.

\begin{figure}	
\includegraphics{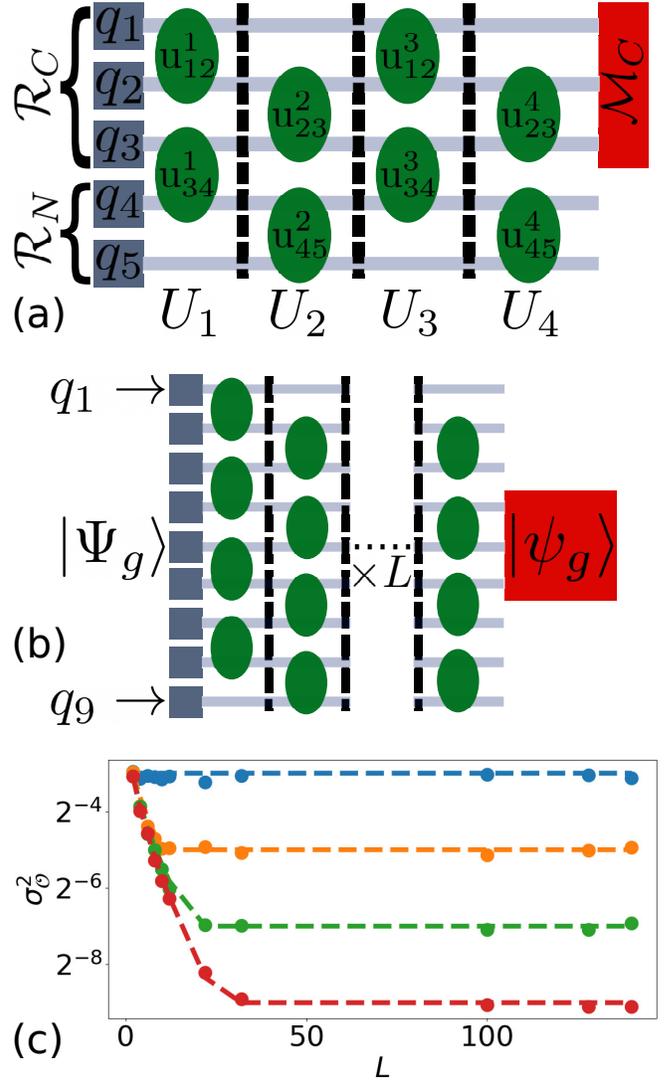}
\caption{\label{fig.Diagram} \textbf{(a)} Diagram of a linear circuit with $n_C = 3$, $n_N = 2$, and $L = 4$. Qubits $q_i$ and $q_j$ interact in layers $k$ through two-qubit unitaries $u_{ij}^k$, which comprise layer unitaries $U_k$, and ultimately form total unitary $U$. The qubits of $\mathcal{R}_C$ are then readout by the cost function operator $\mathcal{M}_C$. \textbf{(b)} Ground state compressor for randomly generated $9$-qubit long-range interaction Hamiltonian (Eq.\ \ref{eq.H}) ground states. The circuit learns to represent the ground states $|\Psi_g\rangle$ as $3$-qubit representations $|\psi_g\rangle$. \textbf{(c)} An illustration of barren plateaus with $\mathcal{L} = \langle \sigma_1^z \sigma_2^z \rangle $ ($n_C = 2$) and $n = 3, 5, 7, 9$ (blue, orange, green, red). The variance $\sigma_\mathcal{O}^2$ of the partial cost function derivative $\mathcal{O}$ is known to decrease rapidly with increasing circuit layers $L$ until ultimately reaching barren plateau magnitude $\sigma_B^2 \propto 2^{-n}$.}
\end{figure}

\section{Variational Algorithms in Layered 1D Quantum Circuits}

Before characterizing the relationship between entanglement and barren plateaus, we provide a brief overview of hybrid quantum-classical variational algorithms in $1$D circuits. Examples of such circuits are shown in Figs.\ \ref{fig.Diagram} (a) and (b). These circuits have a total number of $n$ qubits partitioned into two registers: the cost function register $\mathcal{R}_C$, whose qubits are measured with some observable $\mathcal{M}_C$, and the non-cost function register $\mathcal{R}_N$, with qubits that are not directly measured. These registers have $n_C$ and $n_N$ qubits, respectively, such that $n = n_C + n_N$. In a $1$D system, the qubits interact only with their nearest neighbors via two-qubit unitaries, denoted $u_{ij}^k$ for interactions between the $i$th and $j$th qubit in layer $k$. As this work considers pure states, each $u_{ij}^k$ can be fully describe with six rotation angles as \cite{Dita_2003}

\begin{equation}
u_{ij}^k = R_{34}(\theta_6)R_{23}(\theta_5)R_{12}(\theta_4)R_{34}(\theta_3)R_{23}(\theta_2)R_{34}(\theta_1),
\end{equation}

\noindent where $R_{ij}(\theta)$ is a sinusoidal rotation matrix on axes $i$ and $j$ that can be expressed as $R_{ij}(\theta) = \exp(-i \theta K_{ij})$. Here, $K_{ij}$ is a Hermitian matrix that is equal to $\pm i$ at elements $ij$ and $ji$ and $0$ elsewhere. The universal nature of this parametrization distinguishes our work from studies that impose a restricted unitary structure \cite{Sharma2020, Wiersema2020} and ensures that, for sufficient depth, our unitaries are random enough to generate barren plateaus \cite{McClean2018, Puchala2017}. We note that the $\theta_i$ which correspond to $u_{n_C, n_C+1}^k$ are especially significant, as they entangle registers $\mathcal{R}_C$ and $\mathcal{R}_N$, and we denote them $\theta_i^E$ when relevant.

These two-qubit interactions are then organized into full layer unitaries

\begin{equation}
U_k = \prod_{m = 0}^{(n-1)/2} u_{q+2m, q+2m+1}^k,
\end{equation}

\noindent where $q$ is the remainder of $k/2$. As all interactions are pairwise, the $u_{ij}^k$ in each single-layer unitary $U_k$ commute.

\noindent We describe the unitary of the full system as 

\begin{equation}
U = \prod_{i=1}^L U_i,
\end{equation}

\noindent with total number of gate layers $L$. Fig.\ \ref{fig.Diagram} (a) illustrates a generic example of such a circuit for $n = 5$, $n_C = 3$ and $L=4$.

In hybrid quantum-classical variational algorithms, circuit training is described by a cost function $\mathcal{L}$, which is some function $f$ of expectation value

\begin{equation}
\langle \mathcal{M}_C \rangle = \langle \psi_{out}| \mathcal{M}_C | \psi_{out} \rangle,
\label{eq:expval}
\end{equation}

\noindent such that $\mathcal{L} = f \left[ \langle \mathcal{M}_C \rangle \right]$ and where $|\psi_{out} \rangle = U |\psi_{in} \rangle$ is the output of the quantum circuit $U$ with input state $|\psi_{in} \rangle$. Unless otherwise specified, we take $|\psi_{in} \rangle = | 0 \rangle$. The classical learning algorithm then minimizes $\mathcal{L}$ by updating the parameters $\theta_i$ through use of the partial derivatives 

\begin{equation}
\mathcal{O}_i = \frac{\partial \mathcal{L}} {\partial \theta_i} =\frac{\partial f} {\partial \langle \mathcal{M}_C \rangle} \frac{\partial \langle \mathcal{M}_C \rangle} {\partial \theta_i}.
\end{equation}

Fig.\ \ref{fig.Diagram} (b) illustrates a specific learning example: a ground state compressor. The ground state compressor is a circuit that takes $n=9$ qubit ground states $|\Psi_i^g \rangle$ and their average $z$-axis magnetization

\begin{equation}
\langle M_i \rangle = \langle \Psi_i^g| \frac{1}{n} \sum_i^n \sigma_i^z |\Psi_i^g \rangle,
\end{equation}

\noindent as training data, where $\sigma_i^b$ is the Pauli operator along axis $b$ acting on qubit $i$. The circuit then learns to compress $|\Psi_i^g \rangle$ into $n_C = 3$ qubit equivalents $|\psi_i^g \rangle$ in the $x$-basis by using their average $x$-axis magnetization

\begin{equation}
\langle m_i \rangle = \langle \psi_i^g | \frac{1}{n_C} \sum_i^{n_C} \sigma_i^x |\psi_i^g \rangle
\end{equation}

\noindent as training labels. Here, we generate $N_g$ different $|\Psi_i^g \rangle$ from randomly parametrized long-range interaction Hamiltonians

\begin{equation}
H = \sum_{i,j = 1}^9 \left(J_{ij}^z \sigma_i^z \sigma_j^z + J_{ij}^x \sigma_i^x \sigma_j^x \right) + \sum_{i=1}^9 (w_i \sigma_i^x + v \sigma^z),
\label{eq.H}
\end{equation}

\noindent where $J^z_{ij}$, $J^x_{ij}$, $w_i$, and $v$ are all random constants. In this case, $\langle \mathcal{M}_C \rangle$ is a series of $\langle m_i \rangle$ and we choose $\mathcal{L}$ as the L1 loss between the training output and labels

\begin{equation}
\mathcal{L}_g = \sum_i^{N_g} | \langle m_i \rangle - \langle M_i \rangle |.
\label{eq.L_g}
\end{equation}

\noindent This circuit is an extension of that used in \cite{Shen2020}. We remark that this task is inherently global, requiring magnetization information from both $\mathcal{R}_C$ qubits $4$-$6$ and $\mathcal{R}_N$ qubits $1$-$3$ and $7$-$9$.

\section{The Effect of Entanglement on Barren Plateaus}

Barren plateaus are a manifestation of concentration of measure \cite{Walters2015}, meaning that they arise from the tendency of high-dimensional, random distributions to cluster about their mean. In a PQC, the measurement expectation value $\langle \mathcal{M}_C \rangle$ of the quantum circuit is determined by parameters $\theta_i$. For random circuit initialization, as the number of these parameters grows, the impact of the individual parameter uncertainties becomes small and, for the vast majority of parameter sets $\theta_i$, $\langle \mathcal{M}_C \rangle$ approaches its mean with very low variance such that $\frac{\partial \langle \mathcal{M}_C \rangle}{\partial \theta_i} \rightarrow 0$. In the interest of building intuition, we can draw an analogy between the collective effects of parameters $\theta_i$ on $\langle \mathcal{M}_C \rangle$ and the behavior of an average of $N$ Gaussian distributions $X = \frac{1}{N} \sum_i^N \mathcal{N}_i(\mu, \sigma^2)$, where $\mathcal{N}_i$ are Gaussian distributions with mean $\mu$ and variance $\sigma^2$. Assuming that all $\mathcal{N}_i$ are independent, the uncertainty of individual $\mathcal{N}_i$ are washed out and $X = \mathcal{N}(\mu, \sigma^2 / N)$. That is, the probability that $X$ deviates from $\mu$ vanishes exponentially in $N$.

We emphasize that both concentration of measure and the barren plateaus that they produce are a product of randomness in large dimensional systems, not large dimensionality alone. For this reason, barren plateaus are typically discussed in the context of random PQCs and quantified in terms of unitary $t$-designs \cite{Renes2004, Dankert2009, Harrow2009}, or probability distributions that approximate the average of polynomial functions of degree $\leq t$. Fig.\ \ref{fig.Diagram} (c) illustrates the characteristic behavior of these features, as detailed in \cite{McClean2018}. As the circuits are randomly parametrized, their statistical behavior is described by the Haar distribution. Assuming that $U$ is at least as random as a quantum $1$-design, the mean of $\mathcal{O}_i$ over the probability distribution of all Haar random unitary matrices $U$ is $\mu_{\mathcal{O}_i} = 0$, and the training dynamics rely solely on the variance of this quantity. For relatively shallow circuit depth $L$, it is known that the unitary approaches a quantum $2$-design \cite{Harrow2009}. As such, the variance of the gradient with respect to this unitary ensemble $\sigma_{\mathcal{O}_i}^2 = \text{var}(\mathcal{O}_i)$ decreases rapidly in $L$, ultimately reaching the steady-state $2$-design value $\sim 2^{-n}$ \cite{McClean2018}. As $n$ becomes large, randomly initialized circuit parameters cease to update and training fails. In what follows, we use omitted subscript $\mathcal{O}$ to refer in general to arbitrary parameters $\theta_i$, using the subscripted version $\mathcal{O}_i$ to specify only when the distinction is relevant. For instance, our numerical data is calculated with $\mathcal{O}_1$.

\begin{figure}	
\includegraphics{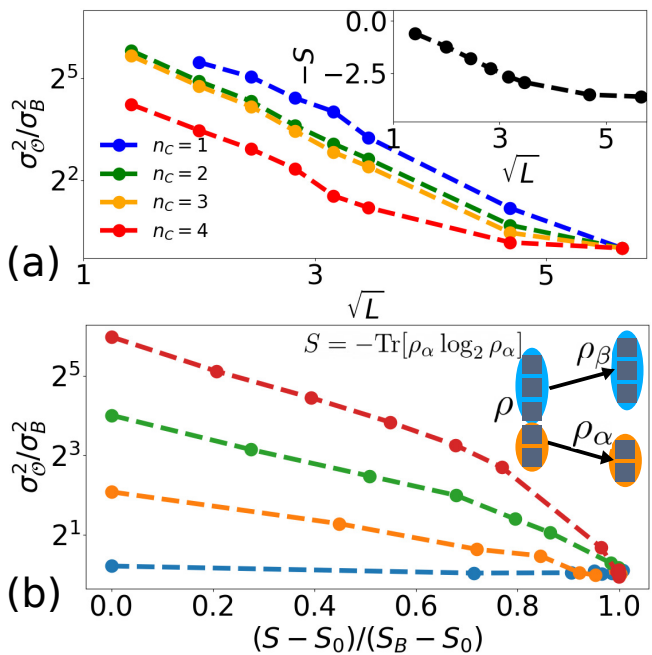}
\caption{\label{fig.BP_vs_L_and_S} \textbf{(a)} Variance $\sigma_\mathcal{O}^2$ of cost function derivative $\mathcal{O}$ for $\mathcal{L} = \langle \prod_{i=1}^{n_C} \sigma_i^z \rangle $ in units of its barren plateau value $\sigma_B^2$ vs number of gate layers $L$ for $n=9$ and various $n_C$. \textbf{(Inset)} The change in entanglement entropy $S$ vs circuit depth $L$ for a $4$-$5$ qubit bipartition, as defined in Eq.\ \ref{eq.S} and illustrated in the inset of (b). Whereas $S$ describes the entanglement growth analytically for $n_C=4$, it is only an approximation for $n_C=1$, $2$, $3$, which could be improved by using an initial $n_C$-$n_N$ bipartition. \textbf{(b)} Variance $\sigma_\mathcal{O}^2$ of cost function derivative $\mathcal{O}$ in units of its barren plateau value $\sigma_B^2$ vs normalized change in entropy $S$ for $\mathcal{L} = \langle \sigma_1^z \sigma_2^z \rangle $ ($n_C = 2$) and $n = 3, 5, 7, 9$ (blue, orange, green, and red). Larger values of $n$ experience greater relative suppression of $\sigma_\mathcal{O}^2$ as $\sigma_\mathcal{O}^2 / \sigma_B^2 \propto S_B = 2^{-S}$, which is the analytical solution for $n=5$ (orange) and an approximation for other $n$. \textbf{(Inset)} Schematic of bi-partition entropy of entanglement $S$ for $n=5$. Full density matrix $\rho$ broken into two subsets $\mathcal{R}_\alpha$ and $\mathcal{R}_\beta$, where $\mathcal{R}_\alpha$ always contains as much of $\mathcal{R}_C$ as possible.}
\end{figure}

To intuitively understand how random entanglement causes barren plateaus, we point out that for a randomly initialized parameter $\theta_i$ to contribute to the concentration of $\langle \mathcal{M}_C \rangle$ and thus to the vanishing of $\mathcal{O}$, it must have some form of influence over the qubits of $\mathcal{R}_C$. For the qubits of $\mathcal{R}_N$, this interaction occurs via $U$ and results in entanglement between the two registers. According to this reasoning, barren plateau emergence should be proportional to the spread of random entanglement. Fig.\ \ref{fig.BP_vs_L_and_S} (a) shows the emergence of barren plateaus vs circuit depth $L$. As $L$ increases, $\sigma_\mathcal{O}^2$ decreases exponentially with $\sqrt{L}$ until approaching its asymptotic limit $\sigma_B^2$. While shallow circuits with smaller cost function registers $\mathcal{R}_C$ initially enjoy greater $\sigma_\mathcal{O}^2$, $n$ determines $\sigma_B^2$ for deep circuits and we will later conjecture that this asymptote corresponds to  entanglement saturation between all qubits on the random circuit. As circuit depth is a form of discretized interaction time $\tau$, this scaling is equivalent to the $\tau$ dependence of entanglement growth of two-level quantum systems in $1$D \cite{Zndaric2020}.

To describe these entanglement dynamics quantitatively, we consider the density matrix of the output qubits

\begin{equation}
\rho =|\psi \rangle \langle \psi | = U |0 \rangle \langle 0 | U^\dagger.
\end{equation}

\noindent In a compromise between simplicity and generality, in this work we describe the spread of circuit entanglement with the bipartite entanglement entropy

\begin{equation}
S = -\text{Tr}[\rho_\alpha \log_2 \rho_\alpha]
\label{eq.S}
\end{equation}

\noindent where $\rho_\alpha$ is the reduced density matrix of $(n-1)/2$ connected qubits of register $\mathcal{R}_\alpha$, taken so as to contain as many cost function qubits as possible. The remaining $(n+1)/2$ qubits are in $\mathcal{R}_\beta$, such that $\rho_\alpha = \text{Tr}_\beta[\rho]$, as illustrated in the inset of Fig.\ \ref{fig.BP_vs_L_and_S} (b). For pure states, this entropy is symmetric and $S = -\text{Tr}[\rho_\beta \log_2 \rho_\beta]$ is equivalent. As this entanglement approximation assumes full entanglement within $\mathcal{R}_\alpha$, it is most precise when $\mathcal{R}_C = \mathcal{R}_\alpha$. Fig.\ \ref{fig.BP_vs_L_and_S} (a) displays $\sigma_\mathcal{O}^2$ vs circuit depth for a variety of $n_C$ in an $n=9$ system. While all $n_C$ scale roughly as $2^{-S}$ (inset), $n_C=4$ is most accurately characterized as, for that case, $\mathcal{R}_C = \mathcal{R}_\alpha$.

Thus, while a single such partitioning is adequate for describing entanglement spread in configurations with $|\mathcal{R}_C| \sim |\mathcal{R}_\alpha| = (n-1)/2$, various such partitions may be used to track short-term entanglement growth when $|\mathcal{R}_C| \ll |\mathcal{R}_\alpha|$ or long-term entanglement growth when $|\mathcal{R}_C| \gg |\mathcal{R}_\alpha|$, such that the entanglement entropy does not temporarily stagnate, like Fig.\ \ref{fig.BP_vs_L_and_S} (b) $n=9$ (red) or rapidly saturate, such as Fig.\ \ref{fig.BP_vs_L_and_S} (b) $n=3$ (blue). The plot is scaled from initial entanglement $S_0$ and normalized to asymptotic difference $S_B - S_0$. In particular, $n=3$ (blue) is initially saturated as it is nearly fully entangled with the minimal number of gates $L=2$, while as $|\mathcal{R}_C| < |\mathcal{R}_\alpha|$, $n=7$, $9$ (green, red) have superlogarithmic scaling for $S \rightarrow S_B$. At the expense of computational simplicity, more general metrics could be adopted, such as a $n_N$-fold sum of bipartite mutual information $I_2$

\begin{equation}
S_N = \sum_{q \in \mathcal{R}_N} I_2(\mathcal{R}_C, \mathcal{R}_q),
\end{equation}

\noindent where $\mathcal{R}_q$ is the single-qubit subspace for each qubit $q \in \mathcal{R}_N$.

We now derive the relationship between $S$ and $\sigma_\mathcal{O}^2$. In particular, we consider $\mathcal{R}_E$, the subspace of qubits that are entangled with (or causal to) the cost function. We begin by proving that $\sigma_\mathcal{O}^2$ is dependent on the dimension $d_E$ of $\mathcal{R}_E$ and then establish the link between $d_E$ and $S$. Let us assume that the circuit input is a product state, here specifically $|0 \rangle$. Then the output state $\rho = |\psi\rangle \langle \psi |$ can be written as $\rho = \rho^E \otimes \rho^D$, where $\rho^E$ belongs to $\mathcal{R}_E$ (is entangled with $\mathcal{M}_C$) and $\rho^D$ does not. This factorization then implies that

\begin{equation}
\rho^E \otimes \rho^D = U| 0 \rangle \langle 0 | U^\dagger \rightarrow U = U_E \otimes U_D,
\label{U_factored}
\end{equation}

\noindent where $U_E|0_E \rangle \langle 0_E| U_E^\dagger = \rho^E$ and $U_D|0_D \rangle \langle 0_D| U_D^\dagger = \rho^D$. Then, the expectation value of our observable $\langle \mathcal{M}_C \rangle$ becomes

\begin{equation}
\langle 0_E | U_E^\dagger \mathcal{M}_C U_E |0_E \rangle \cdot \langle 0_D | U_D^\dagger U_D |0_D \rangle = \langle 0_E | U_E^\dagger \mathcal{M}_C U_E |0_E \rangle,
\end{equation}

\noindent rendering its derivative $\frac{\partial \langle \mathcal{M}_C \rangle}{\partial \theta_i}$ for $\theta_i$ in layer $l$

\begin{equation}
\frac{\partial \langle \mathcal{M}_C \rangle}{\partial \theta_i} = i \langle 0_E| U_R^\dagger \left[K, U_L^\dagger \mathcal{M}_C U_L \right] U_R |0_E \rangle,
\end{equation}

\noindent where $U_R$ and $U_L$ are the products of unitaries $U_k$ for $k < l$ and $k \geq l$, respectively and where $K$ is the rotation generator for $\theta_i$. Assuming a randomly initialized circuit, this reduces the problem to that of \cite{McClean2018}, where using the Haar measure it is shown that $\mu_{\mathcal{O}_i} = 0$ with respect to any $\theta_i$, with variance $\sigma_\mathcal{O}^2 \sim 1 / d_E$, where $d_E = 2^{n_E}$ is the dimensionality of the entangled subspace $\rho^E$. We contrast this with the barrenness of a fully entangled circuit $\sim 1/d = 2^{-(n_E + n_D)} = 2^{-n}$, which can, for many applications, be numerous orders of magnitude smaller.

To implicate $S$ in this barren plateau process, we note that for $\rho = \rho^E \otimes \rho^D$

\begin{equation}
S = -\text{Tr}\left[\text{Tr}_\beta[\rho^E \otimes \rho^D] \cdot \log_2(\text{Tr}_\beta[\rho^E \otimes \rho^D]) \right].
\end{equation}

\noindent Given that if $|\rho^E| < |\rho_\alpha|$ we need to describe early entanglement spread with a smaller bipartition of $S$, we can assume that $\rho^D$ is fully contained in $\rho_\beta$ such that $\text{Tr}_\beta[\rho^E \otimes \rho^D] = \text{Tr}_{\beta_E}[\rho^E] \cdot \text{Tr}[\rho^D] = \text{Tr}_{\beta_E}[\rho^E]$, where $\beta_E$ is the entangled portion of $\mathcal{R}_\beta$. Then

\begin{multline}
S = -\text{Tr}\left[\text{Tr}_{\beta_E}[\rho^E] \cdot \log_2\text{Tr}_{\beta_E}[\rho^E] \right] \\ = -\text{Tr}\left[(\rho^E)_\alpha \cdot \log_2(\rho^{E})_\alpha \right] = n_{\beta_E} = n_E - n_\alpha
\end{multline}

\noindent as this is precisely the definition of the number of entangled qubits shared between $\rho_\alpha$ and $\rho_\beta$. Then, the total number of $\mathcal{R}_C$ entangled qubits is $n_{\beta_E} + n_\alpha = n_E$ such that $d_E = 2^{n_E}$. Therefore, $\sigma_\mathcal{O}^2 \propto 2^{-n_E}$ and changes proportionally to $2^{-S}$.

For simplicity, in the above proofs we assumed that a given qubit is either completely entangled or disentangled. A similar result for the more general case of partial entanglement follows straight-forwardly by taking general $|\psi \rangle = \sum_i c_i |\psi_E^i \rangle |\psi_D^i \rangle$, following the above steps, and repartitioning each component of the sum

\begin{equation}
\rho = \sum_{i,j} c_i c_j^* |\psi_E^i \rangle |\psi_D^i \rangle \langle \psi_E^j | \langle \psi_D^j |
\end{equation}

\noindent into $|\psi_{E'}^k \rangle \langle \psi_{E'}^k | \otimes |\psi_{D'}^k \rangle \langle \psi_{D'}^k |$ such that the dimension of $|\psi_{E'}^k \rangle$, is maximal under the factorization constraint.

\begin{figure}	
\includegraphics{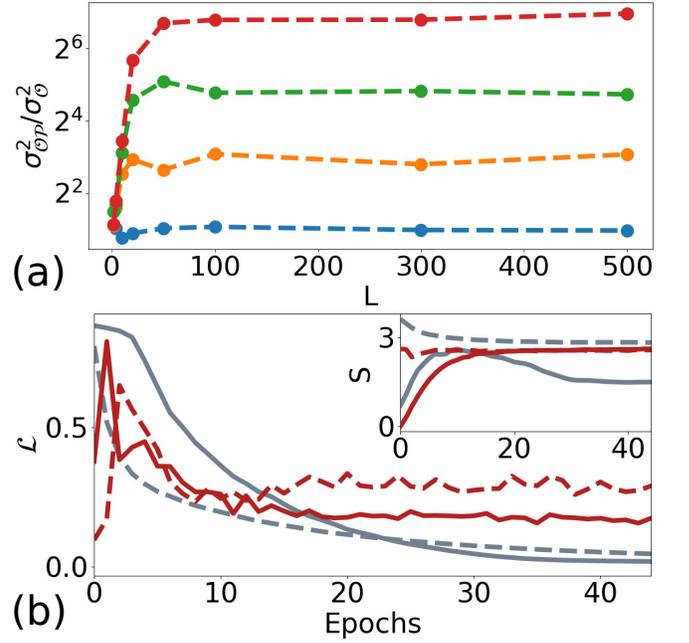}
\caption{\label{fig.Partition_Initial} \textbf{(a)} Partitioned cost function derivative variance $\sigma_\mathcal{OP}^2$ for $\mathcal{L} = \langle \sigma_1^z \sigma_2^z \rangle$ in units of its non-partitioned value $\sigma_\mathcal{O}^2$ vs number of gate layers $L$ for $n = 3, 5, 7, 9$ (blue, orange, green, and red) and $n_C=2$. As expected, $\sigma_\mathcal{OP}^2$ is approximately a factor of $2^{n_N}$ larger than $\sigma_\mathcal{O}^2$ due to the absence of random $\mathcal{R}_C$-$\mathcal{R}_N$ entanglement. \textbf{(b)} Training loss $\mathcal{L} = \mathcal{L}_g$ of Eq.\ \ref{eq.L_g} for ground state compressor in Fig.\ \ref{fig.Diagram} (b) (gray) and $\mathcal{L} = |\langle \sigma_1^z \sigma_2^z \sigma_3^z \rangle |$ (red) vs training epochs for $L=200$. Solid lines are initially partitioned circuits and dashed lines are fully random initializations, with increased learning performance in the latter. The corresponding evolution of $S$ is shown in the (\textbf{inset}).}
\end{figure}

This mapping between the number and degree of cost function-entangled qubits and plateau barrenness highlights that circuit connectivity, and not simply overall circuit depth, is an accurate indicator for the barrenness of the training landscape. Fig.\ \ref{fig.Partition_Initial} (a) is a proof of principle illustration of this point. We remove the register connecting the $u_{2,3}^k$ (unitaries between qubits $q_2$ and $q_3$) gates from each layer $k$ for circuits where $n_C=2$ and $\mathcal{L} = \langle \sigma_1^z \sigma_2^z \rangle$, permanently separating, or partitioning, the registers $\mathcal{R}_C$ and $\mathcal{R}_N$. As circuit depth grows, entanglement with the qubits of $\mathcal{R}_N$ suppresses $\sigma_\mathcal{O}^2$ much faster than its partitioned counterpart $\sigma_\mathcal{OP}^2$, which never exceeds the variance of circuit of total $n = 2$ and is therefore numerous orders of magnitude larger than the variance $\sigma_\mathcal{O}^2$ of the fully entangled system. While insightful, permanent partitioning is clearly not a practical solution for barren plateaus, as it limits not only the barrenness, but also the \textit{expressibility} of the circuit to that of only $n_C$ qubits.

\section{Initialization Techniques for Barren Plateau Mitigation}
\label{sec.initial}

While permanent partitioning is tantamount to simply employing a circuit of smaller $n$, initial parameter restrictions can improve circuit trainability without reducing circuit expressibility. Intuitively, the advantages of this method stem from the role of entropy as a thermodynamic arrow that drives statistical processes forward. As a toy example, let us imagine a classical machine learning protocol where we would like to create a gaseous mixture with optimized concentrations of two gases. If we initially partition the gases, the learning algorithm can simply allow the gases to mix themselves by passing through a vent in the partition, sealing the vent when the ideal concentration is reached on one side. This process occurs independently, driven forward by entropic considerations. If, however, the gases are initially mixed and therefore have a maximum entropy configuration, the learning algorithm cannot succeed by simply unsealing a vent. The problem has been complicated and learning will fail unless more heroic measures are taken.

In this section, we explore methods for quantum equivalents of such entropy limiting initialization schemes and in Sec.\ \ref{sec.dynamic} we detail some of these ``more heroic" measures.

\subsection{Initial Entanglement Partitioning}

One way to avoid barren plateaus without suppressing expressibility is to initially partition the circuit, like in Fig.\ \ref{fig.Partition_Initial} (a), but then to allow $\mathcal{R}_C$-$\mathcal{R}_N$ entanglement throughout the training process. This method is fundamentally distinct from \cite{Grant2019, Wiersema2020}, as we only initialize a subset of two-qubit gates $u_{ij}^k$ to the identity and have devised a cost function and entanglement-based strategy to motivate this choice. Furthermore, our treatment applies to universal PQCs of potentially great depth, not restricted subspaces of $U$ \cite{Wiersema2020}. Fig.\ \ref{fig.Partition_Initial} (b) displays the $n=9$ ground state compressor loss $\mathcal{L} = \mathcal{L}_g$ of Eq.\ \ref{eq.L_g} (gray) and training of $\mathcal{L} = |\langle \sigma_1^z \sigma_2^z \sigma_3^z \rangle |$ (red) for both initially partitioned (solid lines) and fully random (dashed) initializations with $L=200$. Throughout this work, the AMSGrad gradient descent algorithm is used for circuit parameter update \cite{Reddi2018}. The corresponding bipartite entanglement entropies $S$ are in the inset.

At first, initially partitioned circuits suffer a bout of decreased accuracy, which corresponds to a period of low yet rapidly growing entanglement $S$ (inset) that is either insufficient to express the target state of interest (gray) or simply lower than those of many of the degenerate solutions (red). Later, initially partitioned circuits can produce lower error and require fewer training epochs, which we hypothesize stems from the responsiveness of the gradient during the initial phase of low entanglement, enabling the system to train unfettered by barren plateaus and driving interactions forward through entropy growth. Furthermore, strategic initializations may be sufficient to avoid barren plateaus throughout training. We emphasize the relationship between plateau barrenness and $S$ (or alternatively, in other works, $n$ and $L$) is a product of circuit parameter randomness of at least a quantum $2$-design and can thus only be assumed for random circuit initializations. 
That is, such randomness cannot generally be assumed throughout the training process, as this represents an inherently structured organizing of circuit parameters.

When initially partitioned cost functions are learned with high accuracy (here, ground state compression, but also in both the partitioned training of $\mathcal{L} = \langle \sigma_1^z \sigma_2^z \sigma_3^z \rangle$ in Fig.\ \ref{fig.regularize} (c)), $S$ peaks towards the end of the rapid training period before dropping down to a lower steady-state value. This indicates that the initially partitioned circuit identifies an appropriate solution that is less entangled with the unmeasured qubits of $\mathcal{R}_N$, a potentially desirable quality as widespread entanglement can lower the coherence time of qubits. Moreover, an extension of this technique could be used to partially factor the cost function registers themselves, resulting generally in fewer required readouts for a given cost function determination and ameliorating the so-called ``measurement problem" \cite{Verteletskyi2020, Gokhale2019, Izmaylov2019}.

We indicate that this ultimate drop in bipartite entanglement is reminiscent of the late-stage decrease in tripartite mutual information noted in \cite{Shen2020}. The two phenomena are not in conflict, however, with the former indicating the disentanglement of measured and unmeasured qubits \textit{after} the necessary information from those qubits had been collected, while the later signals that the global features of the \textit{input information} are learned towards the end of the training process. In fact, both observations suggest that information locality is the salient feature of late-stage hybrid quantum-classical learning algorithms. Finally, we indicate that a reduction in $S$ also occurs for non-partitioned circuits that can be learned with high accuracy (dashed gray in Fig.\ \ref{fig.Partition_Initial} (b) and dashed black Fig.\ \ref{fig.regularize} (c)), whereas it is absent from low-accuracy circuits (red). Indeed, some degree of automatic $\mathcal{R}_C$-$\mathcal{R}_N$ factorization appears to be a natural feature of high-accuracy QNN training.

Finally, we comment that the cost function $\mathcal{L} = \langle \sigma_1^z \sigma_2^z \sigma_3^z \rangle$ trains rapidly when initialized to a barren plateau, and neither its time nor accuracy are improved by mitigating these barren plateaus via initial partitioning. This suggests that certain classes of cost functions (in this case observables that target one of their eigenstates, as will be discussed in Sec.\ \ref{sec.dynamic}) may be naturally resistant to barren plateaus. This could potentially be due to a rapidly accelerating ordering process, wherein even small $\sigma_\mathcal{O}^2$ quickly navigate $\mathcal{L}$ to a non-barren region of its landscape.

\subsection{Entanglement Meta-Learning as Circuit Pre-Training}

\begin{figure}	
\includegraphics{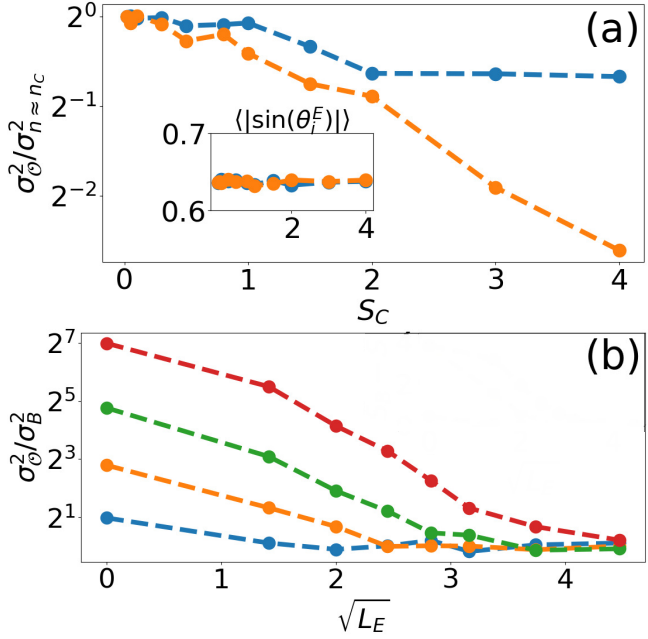}
\caption{\label{fig.limit_pretrain}  \textbf{(a)} A deep circuit ($L=100$) pre-training procedure that minimizes collective entanglement $S_C$ (Eq.\ \ref{eq.SC}), the entanglement entropy between both the input and output registers of $\mathcal{R}_C$ and $\mathcal{R}_N$ for $n=3$, $5$ (blue, orange). As random entanglement decreases with $S_C$, $\sigma_\mathcal{O}^2$ increases. Crucially, this pre-training procedure is unique from partitioned initialization as it permits non-trivial interaction at the level of individual circuit layers \textbf{(inset)}, with magnitude of inter-register interaction remaining $2/\pi \approx 0.637$, which is consistent with that of random $\theta_i$. \textbf{(b)} Derivative variance $\sigma_\mathcal{O}^2$ for $\mathcal{L} = \langle \sigma_1^z \sigma_2^z \rangle$ ($n_C=2$) and initially partitioned registers $\mathcal{R}_C$ and $\mathcal{R}_N$ in units of its non-partitioned value $\sigma_B^2$ vs number of register entangling gate layers $L_E$. Here, $L=200$ and $n = 3, 5, 7, 9$ (blue, orange, green, and red).}
\end{figure}

A similar yet more sophisticated solution for non-barren initializations is classical pre-training of the circuit gates to control $\mathcal{R}_C$-$\mathcal{R}_N$ entanglement. This process is a form of meta-learning \cite{Thrun1998}, a branch of machine learning algorithms directed at optimizing the learning process of other algorithms.

We must be careful, however, in our choice of pre-training cost function. Simply minimizing $S$ would itself be a form of randomly parametrized gradient descent algorithm on the output qubits and would thus, like previous meta-learning techniques, tend to generate the concentration of parameters that lead to barren plateaus \cite{Verdon2019}. We reiterate that this observation is not in conflict with our claim that $\sigma_\mathcal{O}^2$ vanishes $\propto 2^{-S}$ for randomly initiated PQCs, as this relation is not universal, but rather applies to circuit unitaries that are Haar distributed, and can therefore only be assumed in random, not pre-trained, circuits.

As an alternative to $S$, we can combat barrenness by minimizing the \textit{collective} entanglement $S_C$ of registers $\mathcal{R}_C$ and $\mathcal{R}_N$, which considers the $2n$-qubit space of both input and output registers. To define $S_C$, we boost into the $2n$-qubit pure state

\begin{equation}
|\Psi\rangle = \sum_{i,j=0}^{2^n-1}
\frac{\langle \psi_i | U | \psi_j \rangle}{\sqrt{2^n}} |\psi_j\rangle |\psi_i\rangle
\label{eq.Psi_U}
\end{equation}

\noindent where $|\psi_i\rangle$ represent some set basis vectors in the $n$-dimensional Hilbert space. We can then define the density matrix operator of the full $2n$ collective qubit system

\begin{equation}
\mathrm{P} = |\Psi\rangle \langle \Psi|.
\label{eq.Rho}
\end{equation}

\noindent Now the reduced density matrix and corresponding entropy of entanglement of $\mathcal{R}_C$ are defined over its $2n_C$ input \textit{and} output qubits as

\begin{equation}
\mathrm{P}_C = \text{Tr}_N[\mathrm{P}], \hspace{0.2cm} S_C = -\text{Tr}[\mathrm{P}_C \log_2 \mathrm{P}_C],
\label{eq.SC}
\end{equation}

\noindent where $\text{Tr}_N$ is a trace over the $2n_N$ input and output qubits of $\mathcal{R}_N$. Fig.\ \ref{fig.limit_pretrain} (a) shows the growth of $\sigma_\mathcal{O}^2$ for $n=3$, $5$ (blue, orange), $n_C = 2$, and $L=100$. As the entanglement is minimized, $\sigma_\mathcal{O}^2$ draws closer to  its partitioned value $\sigma_{n \approx n_C}^2$, reducing the initialization problem of the barren plateau from $O(2^{-n})$ to approximately $O(2^{-n_C})$ as the initialization $S_C$ of the circuit is minimized through training. This comparison is approximate because the pre-training method, while quite general, implies some inherent ordering that may distinguish it from a bipartition of $2$-designs. As $S_C \rightarrow 0$ and $\mathcal{R}_C$ and $\mathcal{R}_N$ become factorized, $\sigma_\mathcal{O}^2$ grows more similar to the variance of an $n_C$ qubit system, reducing the barren plateau effect by $\approx 2^{n-n_C}$ orders of magnitude. We note that as $S_C \leq 2 \text{min}(n_C, n_N)$, $\sigma_\mathcal{O}^2$ of $n=3$ is constant for $S_C > 2$.

Critically, the average magnitude of $\mathcal{R}_C$-$\mathcal{R}_N$ interaction on a given layer $k$ is not reduced. To see this, consider a rough metric  of inter-register mixing

\begin{equation}
\langle |\sin(\theta_i^E)| \rangle = \frac{1}{3L} \sum_{\theta_i^E} |\sin(\theta_i^E)|,
\end{equation}

\noindent where $\theta_i^E$ are the $3L$ rotation angles of the $u_{n_C, n_C+1}^k$ gates that entangle the registers $\mathcal{R}_C$ and $\mathcal{R}_N$. $\langle |\sin(\theta_i^E)| \rangle$ describes these interactions because it is the average of off-diagonal (or rotating) elements in the two-qubit rotation matrices $u_{n_C, n_C+1}^k$. The inset of Fig.\ \ref{fig.limit_pretrain} demonstrates that this quantity remains at its uniformly distributed value $2 / \pi$, even as the collective registers become increasingly factored. This indicates that while the total entanglement of collective $\mathcal{R}_C$ and $\mathcal{R}_N$ is reduced, the average inter-register interaction at any given layer $k$ remains unaffected, providing a highly non-trivial circuit initialization. This method is distinct from \cite{Grant2019} as it does not produce a network of identity-producing blocks, but rather a nearly arbitrary initialization with the sole yet crucial constraint of adjustable factorizability along a single connection. This distinction may be particularly important for deep circuits \cite{Campos2020}. What is more, this method assumes no specification of problem structure \cite{Verdon2019}, making it generally applicable.

Although classical pre-training is untenable for circuits of large $n$, it can serve as meta-learning for the fundamental effect of entanglement on PQCs and their optimal initializations. As such, it may lead to some scalable generalization of the procedure. It could also be applied iteratively on subsets of large circuits, i.e., on the qubits which form the border between $\mathcal{R}_C$ and $\mathcal{R}_N$. Most promisingly, recent advances in efficient subsystem entanglement measuring techniques, such as random measurements \cite{Brydges260} and fidelity out-of-time correlators \cite{Lewis-Swan2019}, may pave the way for on-hardware hybrid quantum-classical variational minimization of collective entanglement $S_C$, or some analogous measure, enabling full-circuit, high-variance gradient initializations for arbitrary size PQCs.

\section{Dynamic Control of Barren Plateaus}
\label{sec.dynamic}

As the difficulties of training the relatively simple cost function $\mathcal{L} = |\langle \sigma_1^z \sigma_2^z \sigma_3^z \rangle |$ for deep circuits in Fig.\ \ref{fig.Partition_Initial} (b) alludes, initialization techniques can be insufficient for complete mitigation of barren plateaus. To combat this, we now propose a variety of methods to directly manage long-term entanglement of the $\mathcal{R}_C$ and $\mathcal{R}_N$ output registers. Returning to the analogy of optimal mixing of a classical bipartite gas in Sec.\ \ref{sec.initial}, this section details quantum analogies to the more ``heroic" methods that we can take to dynamically control the gaseous mixture's entropy, such as regularization of (penalization of the learning algorithm) or the introduction of additional dynamics into the system.

\subsection{Hard Limit on $\mathcal{R}_C$-$\mathcal{R}_N$ Entangling Gates}

The simplest of these dynamic methods is imposing a hard limit on the number of $\mathcal{R}_C$-$\mathcal{R}_N$ entangling layers $L_E$ in an otherwise deep circuit. This is explored in Fig.\ \ref{fig.limit_pretrain} (b). Although total depth $L=200$, relatively large gradients can still be achieved while still permitting a considerable number of $\mathcal{R}_C$-$\mathcal{R}_N$ interactions. As $L_E$ grows, $\sigma_\mathcal{O}^2$ decays with a similar scaling in $L_E$ as unrestricted circuits do in total gate number $L$, corroborating that barren plateaus indeed arise with the spread of cost function entanglement, not circuit depth itself. This method could be particularly fruitful when using a reinforcement learning algorithm \cite{Zhang2019, Verdon2019}, as the circuit could learn to process and extract the most relevant portions of $\mathcal{R}_N$ before ultimately transferring them to $\mathcal{R}_C$ in a limited number of $L_E$.

\subsection{Entanglement Regularization}

\begin{figure}	
\includegraphics{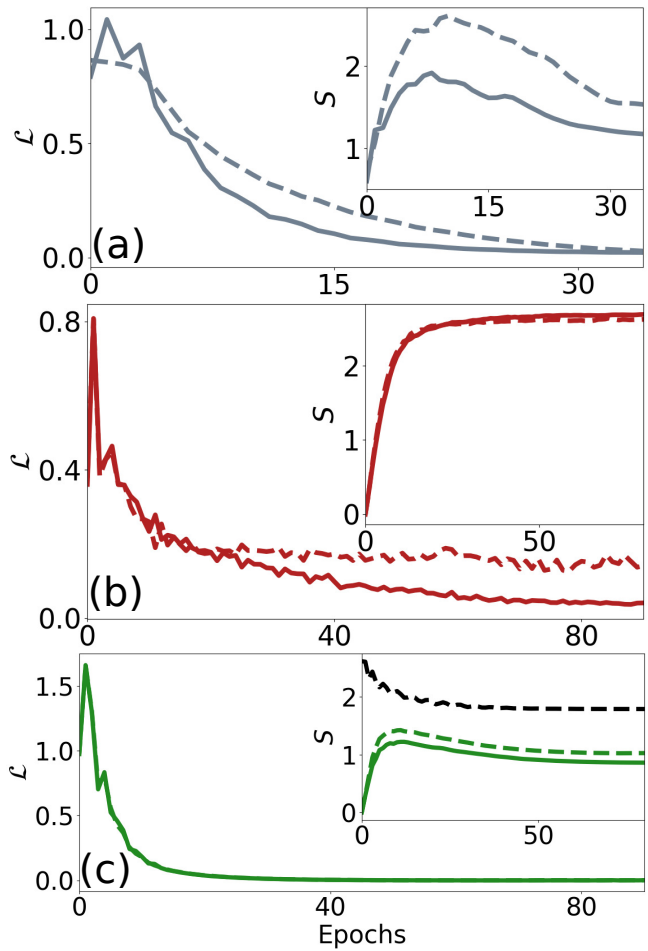}
\caption{\label{fig.regularize} Loss function vs epochs for entanglement regularized learning of \textbf{(a)} ground state compressor $\mathcal{L} = \mathcal{L}_g$ (Eq.\ \ref{eq.L_g}), \textbf{(b)} $\mathcal{L} = |\langle \sigma_1^z \sigma_2^z \sigma_3^z \rangle |$, and \textbf{(c)} $\mathcal{L} = \langle \sigma_1^z \sigma_2^z \sigma_3^z \rangle$ with $L=200$. In all cases, non-zero regularization terms lead to equal \textbf{(c)} or faster and more accurate \textbf{(a and b)} learning with decreased $S$, mitigating the effects of barren plateaus. Moreover, the behavior of $S$ is reflective of the overall training difficulty \textbf{(insets)}: whereas the cost functions of \textbf{(a)} and \textbf{(c)} can be learned rather rapidly and accurately and do so with naturally lower levels of bipartite entanglement $S$ that are more responsive to regularization, that of \textbf{(b)} trains slower and with less accuracy, with entanglement growth that is controlled very little by regularization. The black dashed line in \textbf{(c)} corresponds to $S$ for unregularized, unpartitioned $\mathcal{L} = \langle \sigma_1^z \sigma_2^z \sigma_3^z \rangle$, which learns with equal effectiveness as its partitioned and regularized counterparts, highlighting the increased learnability eigenstate learning.}
\end{figure}

A yet more dynamic method for limiting entanglement is with regularization of $\mathcal{R}_C$-$\mathcal{R}_N$ gates. Regularization adds a penalizing term with adjustable scale parameter $\lambda$

\begin{equation}
\eta = \lambda \sum_i |\sin(\theta_i^E)| \mathcal{L}
\end{equation}

\noindent to the original cost function $\mathcal{L}$ in order to implicitly limit the amount of cross-register entanglement. $\sum_i |\sin(\theta_i^E)|$ is proportional to the inter-register mixing measure $\langle |\sin(\theta_i^E)| \rangle$ of Sec.\ \ref{sec.initial} and serves to limit entanglement generating interactions. Moreover as the values $\theta_i$ are already stored within the classical learning algorithm, this metric does not require additional queries to the quantum hardware. Scaling $\eta$ by $ \mathcal{L} $ results in an adaptive regularization process that resists entanglement in regions of poor solutions while relaxing to the original learning problem as $\mathcal{L}$ approaches zero. This adaptivity can be even more fruitful by making $\lambda$ a decreasing function of $\mathcal{L}$, such that $\eta$ disturbs the learning process even less near optimal solutions.

The regularized gradient is then

\begin{equation}
\mathcal{O}_i = \left(1 + \lambda \sum_i |\sin(\theta_i^E)| \right) \frac{\partial \mathcal{L}}{\partial \theta_i} + \lambda \cos(\theta_i) \text{sign}(\sin(\theta_i)) \mathcal{L}.
\end{equation}

\noindent During portions of the training process that are still largely random, the average of $\frac{\partial \mathcal{L}}{\partial \theta_i}$ over all Haar random unitaries $\mu_{\mathcal{O}_i} = 0$,  as we can assume by concentration of measure for deep circuits that $\sum_i |\sin(\theta_i^E)|$ is approximately constant. The variance, however, does increases to

\begin{equation}
\sigma_{\mathcal{O}_i}^2 \rightarrow \left(1 + \lambda \frac{6L}{\pi} \right)^2 \sigma_{\mathcal{O}_i}^2.
\end{equation}

\noindent We highlight that although regularization only \textit{directly} augments the variance of parameters $\theta_i^E$ upon which it acts, we observe a similar increase on the unregularized angles, indicating that its mitigation of barren plateaus is a system-wide effect. Furthermore, we note that $\lambda$ is adjustable and that the regularized variance grows quadratically in circuit depth, whereas $\sigma_\mathcal{O}^2$ is constant for deep circuits with a given number of qubits $n$.

Fig.\ \ref{fig.regularize} displays this learning process for an initially partitioned circuit trained with a $\lambda$ which is piecewise-adaptive in (solid line) in comparison with an algorithm using only initial partitioning, that is, $\lambda=0$ (dashed) for $L=200$. Three different loss functions are used: (a - gray) ground state compressor $\mathcal{L} = \mathcal{L}_g$ (Eq.\ \ref{eq.L_g}), (b - red) $\mathcal{L} = |\langle \sigma_1^z \sigma_2^z \sigma_3^z \rangle |$, and (c -green) $\mathcal{L} = \langle \sigma_1^z \sigma_2^z \sigma_3^z \rangle$. Ground state compression can be achieved both faster and with greater factorization of the output solution, while solutions to $\mathcal{L} = |\langle \sigma_1^z \sigma_2^z \sigma_3^z \rangle |$ have greatly improved accuracy. Although $\mathcal{L} = \langle \sigma_1^z \sigma_2^z \sigma_3^z \rangle$ is rapidly learned both with and without regularization and/or initial partitioning, its regularized solutions still benefit from the increased factorizability, with the dashed black line in Fig.\ \ref{fig.regularize} (c - inset) showing its unpartitioned $S$.

As discussed in Sec.\ \ref{sec.initial}, we again comment on the seeming resilience from barren plateaus of $\mathcal{L} = \langle \sigma_1^z \sigma_2^z \sigma_3^z \rangle$ and other cost function measurements that target their eigenstates. While still beginning in a barren landscape with overwhelming probability for random circuit initializations, these algorithms learn equally well as without barren plateau mitigation.

\subsection{Langevin Noise as Gradient Supplement}

\begin{figure}	
\includegraphics{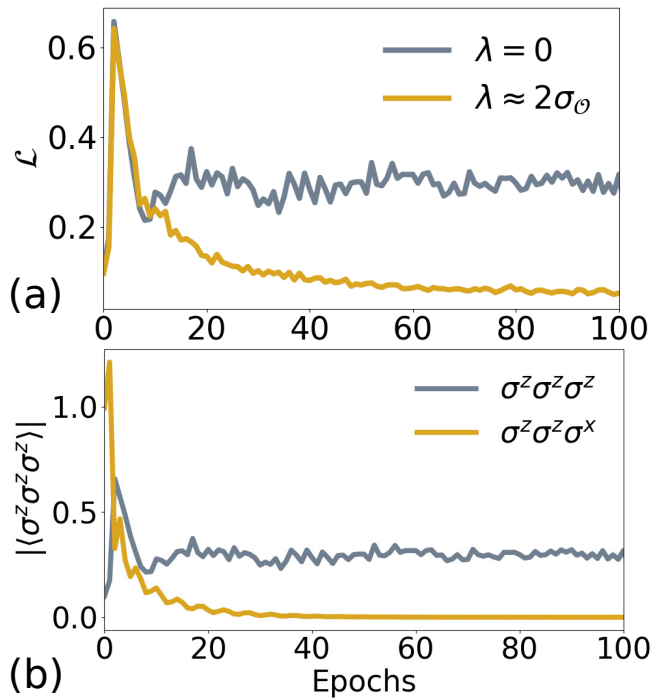}
\caption{\label{fig.basis_langevin} Preparing a state $|\psi\rangle$ under barren plateau conditions (random initialization of $L=200$) for $n=9$ and cost function $\mathcal{L} = |\langle \sigma_1^z \sigma_2^z \sigma_3^z \rangle |$ through both \textbf{(a)} the addition of Langevin noise on a subset of parameters and \textbf{(b)} substitution of measurement basis for which target state is an eigenstate. \textbf{(a)} Additional Langevin noise term $\lambda \sum_i^N |\phi_i| \mathcal{L}$ increases gradient variance $\sigma_{\mathcal{O}_i}^2 \rightarrow (1 + \lambda N \pi)^2 \sigma_{\mathcal{O}_i}^2$ with respect to parameters $\phi_i$, thus helping to navigate barren plateau landscapes. This can be viewed as an increase in diffusion constant $2\mathcal{D}$. \textbf{(b)} By substituting the cost function $\langle \sigma_1^z \sigma_2^z \sigma_3^x \rangle$, for which our target state $|\psi \rangle$ is an eigenstate (or alternatively, choosing a ``natural" cost function basis), we obtain a faster, more accurate learning process.}
\end{figure}

The results of Fig.\ \ref{fig.regularize} (b) raise an interesting point: the addition of regularization terms to the cost function can improve accuracy without significantly decreasing entanglement. We hypothesize that this is because $\eta$, while sometimes successfully limiting entanglement, is always providing additional perturbation in the form of noise. Langevin noise in particular has proven fruitful in classical machine learning and has been used to prevent overfitting in classical neural networks \cite{Welling2011}.

To motivate this hypothesis, we make the observation that the cost function gradient in barren plateaus can be conceptualized as a form of Langevin noise in circuit parameter space. Typically, Langevin noise is defined for functions $g$ that vary with time $\tau$. Then $g(\tau)$ satisfies the conditions that $\langle g(\tau) \rangle_\textbf{Lan} \equiv \int g(\tau) d\tau = 0$ and $\langle g(\tau) g(\tau') \rangle_\textbf{Lan} = 2 \mathcal{D} \delta(\tau-\tau')$, where $\delta$ is the Dirac-delta function and $\mathcal{D}$ is some finite, non-zero diffusion constant \cite{Coffey1996}. In the case of $\mathcal{O}$, the moments are not integrals over time, but rather over the parameters $\theta_i$ as described by the Haar measure, such that $\mathcal{D} = \sigma_\mathcal{O}^2 / 2$.

Under this Langevin noise formulation, barren plateaus can be framed as an entanglement-induced diminution of $\mathcal{D}$. To examine the utility of Langevin noise, we can add an additional noise term $\lambda \sum_i^N |\phi_i| \mathcal{L}$ to the original loss function such that

\begin{equation}
G = (1 + \lambda \sum_i^N |\phi_i| )\mathcal{L},
\end{equation}

\noindent with derivatives $g_i = \frac{\partial G}{\partial \phi_i}$ and where $\phi_i$ are an arbitrarily chosen subset of circuit parameters of size $N$. For uniformly distributed $\phi_i \in \hat{\phi}$ on the interval $[0, 2\pi)$, this yields the equivalent relation

\begin{equation}
\langle g_i(\hat{\phi}) g_i(\hat{\phi}) \rangle_{\textbf{Lan}\mathbf{\phi}} \equiv 2\mathcal{D} = (1 + \lambda N \pi)^2 \sigma_{\mathcal{O}_i}^2.
\end{equation}

\noindent Fig.\ \ref{fig.basis_langevin} (a) illustrates the effectiveness of such Langevin noise in barren landscapes, producing a high-accuracy solution for $\mathcal{L} = | \langle \sigma_1^z \sigma_2^z \sigma_3^z \rangle |$ despite fully random initialization for $n=9$ on a deep circuit ($L=200$). We note that, like the increased variance of entanglement regularization, angles that are not directly perturbed by added Langevin noise still enjoy an increase in variance from the system-wide effect of the technique.

\subsection{Natural Cost Function Bases}
\label{sec.basis}

Finally, we discuss our repeated observation that successful learning in initially barren landscapes is greatly facilitated when the target output is an eigenstate of the cost function observables, a basis choice that we refer to as ``natural". This observation of a natural basis has been made for other product state PQC objective functions, such as in the basis transformations of electronic structure reference states in quantum chemistry \cite{McClean_2016}. Not only do such configurations learn more rapidly, their training rate and accuracy are not impacted by otherwise successful barren plateau mitigation techniques.

We have suggested a potential link between this trainability of natural basis cost functions and the tendency of these circuits to limit their own entanglement, navigating out of the barren landscapes of random matrices and into a tractable configuration. In particular, the variational preparation of measurement eigenstates has several entropy-based advantages, such as a vanishing gradient variance when circuit approaches an optimal solution. As discussed in Sec.\ \ref{sec.initial}, this effect may also be due to a rapidly accelerating ordering process, wherein even small $\sigma_\mathcal{O}^2$ lead to $\mathcal{L}$ efficiently escaping into non-barren regions of its landscape.

Regardless of the origin of its effect, rotating cost function measurements into natural bases can be a strong barren plateau mitigating strategy. Fig.\ \ref{fig.basis_langevin} (b) illustrates that by substituting $\mathcal{L} = \langle \sigma_1^z \sigma_2^z \sigma_3^x \rangle$ for $\mathcal{L} =  | \langle \sigma_1^z \sigma_2^z \sigma_3^z \rangle |$, we can obtain a desired solution $|\psi \rangle$ such that $\langle \psi | \sigma_1^z \sigma_2^z \sigma_3^z |\psi \rangle = 0$ much more effectively. The replacement of a single $\sigma^z$ with the operator $\sigma^x$ reduces the problem to an eigenstate optimization and results in a learning process that trains quickly and automatically limits entanglement, (entanglement behavior analogous to the black dashed line in Fig.\ \ref{fig.regularize} (c)), in contrast to the difficulties of the original problem (red dashed line in Fig.\ \ref{fig.Partition_Initial} (b)).

\section{Conclusion}

We have demonstrated the relationship between total qubit-cost function random entanglement and the barrenness of a learning landscape both analytically and numerically and oriented these findings within the context of many-body entanglement dynamics. Based on these results, we established various metrics for barren plateau prediction, both in terms of entanglement and, for a $1$D system, circuit depth. We also proposed an input-output entanglement metric, whose minimization we suggest is key to circuit learnability. Using this knowledge, we went on to propose various mitigation schemes, including initial partitioning of cost function and non-cost function registers, meta-learning of low-entanglement high-interaction PQC initializations, limiting inter-register interaction, entanglement regularization, the addition of Langevin noise, and utilizing natural cost function bases. We demonstrated the effectiveness of these techniques, elucidating the role that entanglement minimization plays in both the assisted and unassisted training of QNNs and emphasizing that, as existing barren plateau proofs assume sufficiently random parametrizations which do not apply under all circumstance, barren plateaus can potentially be avoided or escaped in generic PQCs.


While these findings imply that QNN learning must strike a non-trivial balance between randomness, expressibility, and barrenness, they lay the groundwork for numerous mitigation techniques that may facilitate large-scale quantum circuit learning. Furthermore, these methods furnish various secondary benefits, such as solution factorization, novel paradigms of quantum meta-learning, and increased understanding of circuit optimization, to name a few. Furthermore, they suggest that the growth of circuit entanglement could potentially be harnessed to drive the learning process.

Oftentimes, the presence of barren plateaus in PQCs is interpreted as an absolute impasse, as it is typically believed to preclude learning. However, this work emphasizes that not only can barren plateaus be ameliorated through entanglement considerations, they should be understood as manifestations of circuit randomness that have not been proved to apply to more organized configurations, such as those which may manifest during the learning process. To understand the relationship between cost function barrenness and total circuit learnability, the evolution of circuit parameter distributions throughout the learning process should be characterized. Such a statistical characterization will also shed further light on the viability of barren plateau mitigation methods.



\begin{acknowledgments}
S.F.Y. and T.L.P. would like to thank the AFOSR and the NSF  for funding through the CUA-PFC grant. T.L.P. acknowledges that this material is based upon work supported by the National Science Foundation Graduate Research Fellowship under Grant No. DGE-1745303. X.G. is supported by the Postdoctoral Fellowship in Quantum Science of the Harvard-MPQ Center for Quantum Optics, the Templeton Religion Trust grant TRT 0159, and by the Army Research Office under Grant W911NF1910302 and MURI Grant W911NF-20-1-0082.
\end{acknowledgments}

\bibliography{MyCollection}

\end{document}